\documentclass[12pt]{article}
\usepackage{latexsym,amsfonts,amssymb}
\makeatletter
\@addtoreset{equation}{subsection}
\makeatother

\begin{document}
\begin{center}
{\Large \bf Gauge Boson Theory of Quantum State Reduction}
\\[1.5cm]
 {\bf Vladimir S.~MASHKEVICH}\footnote {E-mail:
  Vladimir.Mashkevich100@qc.cuny.edu}
\\[1.4cm] {\it Physics Department
 \\ Queens College\\ The City University of New York\\
 65-30 Kissena Boulevard\\ Flushing\\ New York
 11367-1519} \\[1.4cm] \vskip 1cm

{\large \bf Abstract}
\end{center}

A theory of quantum state reduction is advanced. It is based on
two principles: (1) Gauge decomposition; (2) Maximum entropy. To
wit: (1) The reduction decomposition of a state vector is the
Schmidt decomposition with respect to the states of a set of
(dressed) gauge boson modes; (2) The reduction instant is that of
the maximum entropy of a resulting mixed state. The theory
determines states undergoing the reduction, its instant, resulting
pure states and their probabilities. Applications: (Polarized)
photon absorption and transmission, emission, particle detection,
reduction of a superposition of states, nonintegral photon states,
photon and matter-photon entanglement, processes with weak bosons,
and the role of gluons.

\newpage

\section*{Introduction and motivation}

Conventionally, the quantum state reduction problem is considered
in the light of the measurement problem [1-9] and reduces to the
problem of reduction of a superposition of (macroscopically
different) states of an apparatus. It appears, however, that it is
the concept of state reduction that is primary rather than that of
measurement. (In fact, this was Bell`s point of view [10,11].)
Thus, state reduction should be treated in its own right.

The challenge of constructing a theory of reduction involves the
problems of determining (1)~ decomposition of states undergoing
reduction, (2) its instant, (3) resulting states and (4)~their
probabilities; the solution to the latter two problems amounts to
that of the former two.

A starting point is this. A dynamical theory of state reduction
should be based on an actual interaction, which is assumed to
involve gauge bosons. Following Dirac [12], consider a photon
passing through a crystal of tourmaline. In the case of the photon
polarized obliquely to the optic axis, unitary time evolution
results in a superposition of the two states: $|\mathrm{no\;
photons}\rangle\otimes|\mathrm{crystal \;excited\rangle}$ and
$|\mathrm{one\;photon\;polarized\;perpendicular\;to\;the\;axis}
\rangle\otimes|\mathrm{crystal\;unchanged\rangle}$. Thus the
photon mode is entangled with the crystal, i.e., the rest of the
system. The state vector has the form of a Schmidt decomposition.
But in an actual experiment, the entanglement does not last: there
occurs a quantum jump, which results in a disentangled state.

In general, we may assume that a gauge interaction gives rise to
an entanglement of a set of gauge boson modes with the rest of the
universe, and  reduction causes the disentanglement. This leads us
to the solution of the decomposition problem.

Had there been no disentanglement, the world would be a complete
mess. It is gauge bosons that engender entanglement, so let them
play a crucial role in disentanglement.

In the simplest case, like the one above, the state of a boson
mode is integral (i.e., with an integral number of bosons), but
nonintegral states are possible as well (for example, a coherent
state of a laser mode).

Now turn to the problem of reduction instant. The reduction of a
pure state results in an increase of entropy. Therefore it is
natural to define the reduction instant as that of the maximum
entropy increase.

\section{Two basic principles}

\subsection{The principle of gauge decomposition}

The gauge decomposition principle determines the form of the
decomposition of a state vector undergoing reduction. It reads:

{\it The reduction decomposition of a state vector is the Schmidt
decomposition with respect to the states of a set of (dressed)
gauge boson modes.}

The decomposition is of the form
\begin{equation}\label{1.1.1}
|\;\rangle=\sum_{j}c_{j}|\mathrm{G}j\rangle\otimes|\mathrm{R}j\rangle:=
\sum_{j}c_{j}|\mathrm{Gauge}j\rangle\otimes|\mathrm{Rest}j\rangle
\end{equation}
\begin{equation}\label{1.1.2}
\langle\mathrm{G}j\,'|\mathrm{G}j\rangle=
\delta_{j\,'j}\qquad\langle\mathrm{R}j\,'|\mathrm{R}j\rangle=\delta_{j\,'j}
\qquad\sum_{j}|c_{j}|^{2}=1
\end{equation}
The reduction of the above pure state at an instant
$t=t_{\mathrm{red}}$ is this:
\begin{equation}\label{1.1.3}
\hat{\rho}(t_{\mathrm{red}}-0)\stackrel{\mathrm{red}}\longrightarrow
\hat{\rho}(t_{\mathrm{red}})
\end{equation}
\begin{equation}\label{1.1.4}
\hat{\rho}(t_{\mathrm{red}}-0)=|\;\rangle\langle\;|
\end{equation}
\begin{equation}\label{1.1.5}
\hat{\rho}(t_{\mathrm{red}})=\sum_{j}w_{j}
[\,|\mathrm{G}j\rangle\langle\mathrm{G}j|\,]\otimes
[\,|\mathrm{R}j\rangle\langle\mathrm{R}j|\,]\qquad\quad
w_{j}=|c_{j}|^{2}
\end{equation}
i.e.,
\begin{equation}\label{1.1.6}
|t_{\mathrm{red}}-0\rangle\stackrel{\mathrm{red}}\longrightarrow|t_{\mathrm{red}}\rangle
\end{equation}
\begin{equation}\label{1.1.7}
|t_{\mathrm{red}}-0\rangle=\sum_{j}c_{j}|\mathrm{G}j\rangle\otimes|\mathrm{R}j\rangle
\end{equation}
\begin{equation}\label{1.1.8}
|t_{\mathrm{red}}\rangle=|\mathrm{G}j\rangle\otimes|\mathrm{R}j\rangle\qquad
\mathrm{with\;probability}\;w_{j}=|c_{j}|^{2}
\end{equation}

The reduction of a mixed state of the form
\begin{equation}\label{1.1.9}
\hat{\rho}(t_{\mathrm{red}}-0)=\sum_{k}p_{k}|k\rangle\langle
k|\qquad \langle k|k\rangle=1\qquad |\langle k\,'|k\rangle|\leq
1\qquad \sum_{k}p_{k}=1
\end{equation}
is determined in the following way. The Schmidt decomposition is
\begin{equation}\label{1.1.10}
|k\rangle=\sum_{j_{k}}c_{kj_{k}}|kj_{k}\rangle\qquad|kj_{k}\rangle=
|\mathrm{G}kj_{k}\rangle\otimes|\mathrm{R}kj_{k}\rangle
\end{equation}
\begin{equation}\label{1.1.11}
\langle\mathrm{G/R}kj\,'_{k}|\mathrm{G/R}kj_{k}\rangle=\delta_{j_{k}\,'j_{k}}
\qquad
|\langle\mathrm{G/R}k\,'j\,'_{k\,'}|\mathrm{G/R}kj_{k}\rangle|\leq
1
\end{equation}
Now
\begin{equation}\label{1.1.12}
\hat{\rho}(t_{\mathrm{red}}-0)\stackrel{\mathrm{red}}\longrightarrow
\hat{\rho}(t_{\mathrm{red}})=\sum_{k}p_{k}\sum_{j_{k}}w_{kj_{k}}|kj_{k}\rangle\langle
kj_{k}|\qquad w_{kj_{k}}=|c_{kj_{k}}|^{2}
\end{equation}
i.e.,
\begin{equation}\label{1.1.13}
\hat{\rho}(t_{\mathrm{red}}-0)\stackrel{\mathrm{red}}\longrightarrow
|kj_{k}\rangle\langle
kj_{k}|\Leftrightarrow|kj_{k}\rangle\qquad\mathrm{with\;probability}\;p_{k}w_{kj_{k}}
\end{equation}
and
\begin{equation}\label{1.1.14}
\sum_{k}\sum_{j_{k}}p_{k}w_{kj_{k}}=\sum_{k}p_{k}=1
\end{equation}

Notice that the result of the reduction is a pure state (1.1.8) or
(1.1.13), respectively.

The representation (1.1.9) of the mixed state is not uniquely
defined; it will be fixed in the following Subsection.

\subsection{The principle of maximum entropy}

The maximum entropy principle determines the reduction instant,
$t_{\mathrm{red}}$, and, by the same token, actual resulting
states and their probabilities. It reads:

{\it The reduction instant is that of the maximum entropy of the
resulting mixed state.}

Introduce
\begin{equation}\label{1.2.1}
\hat{\rho}_{\mathrm{red}}(t):=\sum_{j}w_{j}|j\rangle\langle
j|\quad \mathrm{or} \quad
\sum_{k}p_{k}\sum_{j_{k}}w_{kj_{k}}|kj_{k}\rangle\langle kj_{k}|
\end{equation}
in the case of the reduction of a pure (1.1.6) or mixed (1.1.13)
state, respectively. The entropy
\begin{equation}\label{1.2.2}
\sigma_{\mathrm{red}}(t)=
-\mathrm{Tr}\{\hat{\rho}_{\mathrm{red}}(t)\ln\hat{\rho}_{\mathrm{red}}(t)\}
\end{equation}

In the pure case
\begin{equation}\label{1.2.3}
\sigma_{\mathrm{red}}(t)=-\sum_{j}w_{j}\ln w_{j}\qquad
w_{j}=w_{j}(t)
\end{equation}
and $t_{\mathrm{red}}$ is determined by
\begin{equation}\label{1.2.4}
\max\limits_t
\,\sigma_\mathrm{red}(t)=\sigma_\mathrm{red}(t_{\mathrm{red}})
\qquad\frac{\displaystyle\mathrm{d}\sigma_\mathrm{red}}{\displaystyle\mathrm{d}t}=0
\end{equation}

In the mixed case, we introduce
\begin{equation}\label{1.2.5}
\sigma_{\mathrm{red\,max}}(t)=\max\limits_{\{|k\rangle\}}\,\sigma_{\mathrm{red}}(t)
\end{equation}
where $\{|k\rangle\}$ is the set of vectors $|k\rangle$ in the
representation of the mixed state (1.1.9); this fixes the
representation. Now $t_{\mathrm{red}}$ is determined by
\begin{equation}\label{1.2.6}
\max\limits_t
\,\sigma_\mathrm{red\,max}(t)=\sigma_\mathrm{red\,max}(t_{\mathrm{red}})
\qquad\frac{\displaystyle\mathrm{d}\sigma_\mathrm{red\,max}}{\displaystyle\mathrm{d}t}=0
\end{equation}

Let us consider the pure case in more detail. We have
\begin{equation}\label{1.2.7}
\frac{\mathrm{d}\sigma_{\mathrm{red}}}{\mathrm{d}t}=
\sum_{j}\frac{\partial\sigma_{\mathrm{red}}}{\partial
w_{j}}\frac{\mathrm{d}w_{j}}{\mathrm{d}t}=\sum_{j}(-\ln
w_{j}-1)\frac{\mathrm{d}w_{j}}{\mathrm{d}t}=-\sum_{j}(\ln
w_{j})\frac{\mathrm{d}w_{j}}{\mathrm{d}t}
\end{equation}
so that
\begin{equation}\label{1.2.8}
\sum_{j}(\ln w_{j})\frac{\mathrm{d}w_{j}}{\mathrm{d}t}=0
\qquad\mathrm{with}\;\sum_{j}\frac{\mathrm{d}w_{j}}{\mathrm{d}t}=0
\end{equation}

Let $j=1,2$. Then
\begin{equation}\label{1.2.9}
\left[\ln\left(\frac{1}{w_{1}}-1\right)\right]\frac{\mathrm{d}w_{1}}{\mathrm{d}t}=0
\end{equation}
whence either
\begin{equation}\label{1.2.10}
w_{2}=w_{1}=\frac{1}{2}
\end{equation}
or
\begin{equation}\label{1.2.11}
\frac{\mathrm{d}w_{2}}{\mathrm{d}t}=\frac{\mathrm{d}w_{1}}{\mathrm{d}t}=0
\end{equation}

\section{Some implications}

\subsection{Cluster noncorrelatedness}

There is ``a crucial physical requirement, the cluster
decomposition principle, which says in effect that distant
experiments yield uncorrelated results''[13]. Let us show that the
principle of maximum entropy provides the noncorrelatedness of
reduction in independent systems.

Let $1$ and $2$ be such systems and
\begin{equation}\label{2.1.1}
|\;\rangle=|1\rangle\otimes|2\rangle
\end{equation}
(for the sake of simplicity, we consider pure states). The Schmidt
decomposition is
\begin{equation}\label{2.1.2}
|l\rangle=\sum_{j_{l}}c_{j_{l}}|\mathrm{G}lj_{l}\rangle
\otimes|\mathrm{R}lj_{l}\rangle,\;l=1,2
\end{equation}
Consider the possibility of reduction of the composite system. We
have
\begin{equation}\label{2.1.3}
|\;\rangle=\sum_{j_{1}j_{2}}c_{j_{1}j_{2}}|\mathrm{G}j_{1}j_{2}
\rangle\otimes|\mathrm{R}j_{1}j_{2}\rangle
\end{equation}
\begin{equation}\label{2.1.4}
|\cdot j_{1}j_{2}\rangle=|1\cdot j_{1}\rangle\otimes|2\cdot
j_{2}\rangle\qquad c_{j_{1}j_{2}}=c_{j_{1}}c_{j_{2}}
\end{equation}
so that
\begin{equation}\label{2.1.5}
|\;\rangle\stackrel{\mathrm{red}}\longrightarrow
|\mathrm{G}j_{1}j_{2}\rangle\otimes|\mathrm{R}j_{1}j_{2}\rangle
\qquad\mathrm{with\;
probability}\;w_{j_{1}j_{2}}=|c_{j_{1}j_{2}}|^2=
|c_{j_{1}}|^2|c_{j_{2}}|^2=:w_{j_{1}}w_{j_{2}}
\end{equation}
Now
\begin{equation}\label{2.1.6}
\hat{\rho}_{\mathrm{red}}=\hat{\rho}_{1\mathrm{red}}
\otimes\hat{\rho}_{2\mathrm{red}}\qquad\hat{\rho}_{l\mathrm{red}}=
\sum_{j_{l}}w_{lj_{l}}|lj_{l}\rangle\langle
lj_{l}|\qquad|lj_{l}\rangle=
|\mathrm{G}lj_{l}\rangle\otimes|\mathrm{R}lj_{l}\rangle\
\end{equation}
Thus
\begin{equation}\label{2.1.7}
\sigma_{\mathrm{red}}=\sigma_{1\mathrm{red}}+\sigma_{2\mathrm{red}}
\end{equation}
From
\begin{equation}\label{2.1.8}
\frac{\mathrm{d}\sigma_{\mathrm{red}}}{\mathrm{d}t}=0\quad\mathrm{and}
\quad\frac{\mathrm{d}\sigma_{l\mathrm{red}}}{\mathrm{d}t} \neq
0,\quad l=1,2,\quad t=t_{\mathrm{red}}
\end{equation}
follows
\begin{equation}\label{2.1.9}
\frac{\mathrm{d}\sigma_{1\mathrm{red}}}{\mathrm{d}t}
\frac{\mathrm{d}\sigma_{2\mathrm{red}}}{\mathrm{d}t}<0
\end{equation}
Let
\begin{equation}\label{2.1.10}
\frac{\mathrm{d}\sigma_{1\mathrm{red}}}{\mathrm{d}t}<0
\qquad\frac{\mathrm{d}\sigma_{2\mathrm{red}}}{\mathrm{d}t}>0\qquad
t=t_{\mathrm{red}}
\end{equation}
Then a reduction in the system $1$ should have occurred at
$t_{1\mathrm{red}}<t_{\mathrm{red}}$ when
\begin{equation}\label{2.1.11}
\frac{\mathrm{d}\sigma_{1\mathrm{red}}}{\mathrm{d}t}=0\qquad
t=t_{\mathrm{red}}
\end{equation}
Thus we have
\begin{equation}\label{2.1.12}
t_{1\mathrm{red}}<t_{\mathrm{red}}<t_{2\mathrm{red}}
\end{equation}
and $t_{\mathrm{red}}$ does not correspond to any reduction.

\subsection{Reduction, nonlocality, and relativity}

Quantum state reduction is a nonlocal phenomenon. As for the
relativistic aspect of the reduction, there are two possible
points of view. On the one hand, the stated theory may be
considered to be nonrelativistic. On the other hand, it is
possible to assume that the reduction occurs in the cosmic
reference frame, so that $t$ (and $t_{\mathrm{red}}$) is cosmic
time. It is quantum jumps that click cosmic time.

\subsection{The role of fluctuations}

In view of possible fluctuations, a state with
$\mathrm{d}\sigma_{\mathrm{red}}/\mathrm{d}t\rightarrow +\,0$ may
be unstable with respect to reduction.

\section{Applications: Integral photon states}

\subsection{Photon passing through a tourmaline crystal}

First of all, let us return to a classic example of reduction: a
photon passing through a crystal of tourmaline. The initial state
is
\begin{equation}\label{3.1.1}
|t=0\rangle=[c_{\bot}|\mathrm{M_{\bot}}1\rangle\otimes
|\mathrm{M}_{\|}0\rangle+c_{\|}|\mathrm{M_{\bot}}0\rangle\otimes
|\mathrm{M}_{\|}1\rangle]\otimes|\mathrm{T}0\rangle\qquad
c_{\bot}=\sin\alpha\quad c_{\|}=\cos\alpha
\end{equation}
where $\mathrm{M}$ stands for photon mode, $\bot/\|$ for
polarization, and $\mathrm{T}$ for tourmaline. A unitary time
evolution is of the form
\begin{eqnarray}
|t=0\rangle\stackrel{\mathrm{U}}{\rightarrow}|t\rangle
=c_{\bot}[\,|\mathrm{M_{\bot}}1\rangle\otimes
|\mathrm{M}_{\|}0\rangle]\otimes|\mathrm{T}0\rangle+
c_{\|}|\mathrm{M}_{\bot}0\rangle\otimes\{\mu_{1}^{1}
|\mathrm{M}_{\|}1\rangle\otimes|\mathrm{T}0\rangle+
\mu_{0}^{1}|\mathrm{M}_{\|}0\rangle\otimes|\mathrm{T}1\rangle\}\nonumber\\
 =[c_{\bot}|\mathrm{M_{\bot}}1\rangle\otimes
|\mathrm{M}_{\|}0\rangle+c_{\|}\mu_{1}^{1}|\mathrm{M_{\bot}}0\rangle\otimes
|\mathrm{M}_{\|}1\rangle]\otimes|\mathrm{T}0\rangle+c_{\|}\mu_{0}^{1}
[\,|\mathrm{M_{\bot}}0\rangle\otimes
|\mathrm{M}_{\|}0\rangle\,]\otimes|\mathrm{T}1\rangle\nonumber\\
\mu=\mu(t)\qquad |\mu_{0}^{1}|^{2}+|\mu_{1}^{1}|^{2}=1\qquad
\end{eqnarray}

Thus for the first reduction,
\begin{equation}\label{3.1.3}
\hat{\rho}_{\mathrm{red}}^{1}(t)=
\sum_{j}^{0,1}w_{j}^{1}|j\rangle\langle j|
\end{equation}
with the states
\begin{equation}\label{3.1.4}
|0\rangle=[\,|\mathrm{M_{\bot}}0\rangle\otimes
|\mathrm{M}_{\|}0\rangle\,]\otimes|\mathrm{T}1\rangle\qquad
|1\rangle=\frac{1}{\sqrt{w_{1}^{1}}}[c_{\bot}|\mathrm{M_{\bot}}1\rangle\otimes
|\mathrm{M}_{\|}0\rangle+c_{\|}\mu_{1}^{1}|\mathrm{M_{\bot}}0\rangle\otimes
|\mathrm{M}_{\|}1\rangle\,]\otimes|\mathrm{T}0\rangle
\end{equation}
and probabilities
\begin{equation}\label{3.1.5}
\mathrm{no\;photons}:\;w_{0}^{1}=|c_{\|}|^{2}|\mu_{0}^{1}|^{2}\qquad
\mathrm{one\;photon}:\;w_{1}^{1}=|c_{\bot}|^{2}+|c_{\|}|^{2}|\mu_{1}^{1}|^{2}
\end{equation}

Now let $t=t_{\mathrm{red}}$ and use (1.2.10), (1.2.11). If
$|c_{_{\|}}|\leq 1/2$, then
\begin{equation}\label{3.1.6}
|\mu_{0}^{1}|=1\qquad \mu_{1}^{1}=0\qquad
w_{0}^{1}=|c_{\|}|^{2}\qquad w_{1}^{1}=|c_{\bot}|^{2}
\end{equation}
\begin{equation}\label{3.1.7}
|1\rangle=[\,|\mathrm{M_{\bot}}1\rangle\otimes
|\mathrm{M}_{\|}0\rangle\,]\otimes|\mathrm{T}0\rangle
\end{equation}
and the first reduction is the only one.

Let now $|c_{\|}|^{2}>1/2$. In this case,
\begin{equation}\label{3.1.8}
w_{0}^{1}=w_{1}^{1}=\frac{1}{2}\qquad
|\mu_{0}^{1}|^{2}=\frac{1}{2|c_{\|}|^{2}}<1
\end{equation}
and the resulting one-photon state is
\begin{equation}\label{3.1.9}
|1\rangle=\sqrt{2}[c_{\bot}|\mathrm{M_{\bot}}1\rangle\otimes
|\mathrm{M}_{\|}0\rangle+c_{\|}\mu_{1}^{1}|\mathrm{M_{\bot}}0\rangle\otimes
|\mathrm{M}_{\|}1\rangle\,]\otimes|\mathrm{T}0\rangle
\end{equation}
Now consider the $\mathrm{U}$ evolution of the one-photon state
(3.1.9). If the resulting state is
\begin{equation}\label{3.1.10}
\sqrt{2}c_{\bot}[\,|\mathrm{M_{\bot}}1\rangle\otimes
|\mathrm{M}_{\|}0\rangle\,]\otimes|\mathrm{T}0\rangle+
(1-2|c_{\bot}|^{2})^{1/2}[\,|\mathrm{M_{\bot}}0\rangle\otimes
|\mathrm{M}_{\|}0\rangle\,]\otimes|\mathrm{T}1\rangle
\end{equation}
so that under the second reduction
\begin{equation}\label{3.1.11}
w_{1}^{2}=2|c_{\bot}|^{2}
\end{equation}
then the second reduction is the last one and the total
probabilities are
\begin{equation}\label{3.1.12}
W_{1}=w_{1}^{1}w_{1}^{2}=\frac{1}{2}\times
2|c_{\bot}|^{2}=|c_{\bot}|^{2}\qquad
W_{0}=1-|c_{\bot}|^{2}=|c_{\|}|^{2}
\end{equation}
Otherwise we proceed in the same way. The final result is this:
\begin{equation}\label{3.1.13}
\frac{1}{2^{n}}\leq|c_{\bot}|^{2}<\frac{1}{2^{n-1}}\qquad
n\;\mathrm{reductions}
\end{equation}
\begin{equation}\label{3.1.14}
 W_{1}=w_{1}^{1}w_{1}^{2}\cdots
w_{1}^{n}=\left(\frac{1}{2}\right)^{n-1}
\left[\left(\sqrt{2}\right)^{n-1}\right]^{2}
|c_{\bot}|^{2}=|c_{\bot}|^{2}\qquad W_{0}=|c_{\|}|^{2}
\end{equation}

\subsection{Absorption and transmission}

Let absorption and transmission factors be $p_{\mathrm{abs}}$ and
$p_{\mathrm{trans}}$, respectively. The initial state is
\begin{equation}\label{3.2.1}
|t=0\rangle=|\mathrm{M}1\rangle\otimes|\mathrm{A}0\rangle
\end{equation}
where $\mathrm{A}$ stands for an absorbing medium. Now
\begin{equation}\label{3.2.2}
|\mathrm{M}1\rangle\otimes|\mathrm{A}0\rangle
\stackrel{\mathrm{U}}{\rightarrow}
\mu_{1}^{1}|\mathrm{M}1\rangle\otimes|\mathrm{A}0\rangle+
\mu_{0}^{1}|\mathrm{M}0\rangle\otimes|\mathrm{A}1\rangle\qquad
|\mu_{1}^{1}|^{2}+|\mu_{0}^{1}|^{2}=1\qquad \mu=\mu(t)
\end{equation}
so that
\begin{equation}\label{3.2.3}
\hat{\rho}_{\mathrm{red}}^{1}(t)=
\sum_{j}^{0,1}w_{j}^{1}|j\rangle\langle j|
\end{equation}
with the states
\begin{equation}\label{3.2.4}
|0\rangle=|\mathrm{M}0\rangle\otimes|\mathrm{A}1\rangle\qquad
|1\rangle=|\mathrm{M}1\rangle\otimes|\mathrm{A}0\rangle
\end{equation}
and probabilities
\begin{equation}\label{3.2.5}
w_{j}^{1}=|\mu_{j}^{1}|^{2}\,,\;j=0,1
\end{equation}
The reduction is determined by (1.2.10), (1.2.11). If
$p_{\mathrm{abs}}\leq1/2$, then
\begin{equation}\label{3.2.6}
(w_{1}^{1})_{\mathrm{max}}=p_{\mathrm{abs}}
\end{equation}
so that there occurs only one reduction, with
\begin{equation}\label{3.2.7}
w_{0}^{1}=p_{\mathrm{abs}}\qquad w_{1}^{1}=p_{\mathrm{trans}}
\end{equation}

If $p_{\mathrm{abs}}>1/2$, then
\begin{equation}\label{3.2.8}
w_{0}^{1}=\frac{1}{2}\qquad w_{1}^{1}=\frac{1}{2}
\end{equation}
and
\begin{equation}\label{3.2.9}
|1\rangle=|\mathrm{M}1\rangle\otimes|\mathrm{A}0\rangle
\stackrel{\mathrm{U}}{\rightarrow}\mu_{1}^{2}
|\mathrm{M}1\rangle\otimes|\mathrm{A}0\rangle+\mu_{0}^{2}
|\mathrm{M}0\rangle\otimes|\mathrm{A}1\rangle
\end{equation}
The final result:
\begin{equation}\label{3.2.10}
n\;\mathrm{reductions}\qquad
W_{1}=\left(\frac{1}{2}\right)^{n-1}|\mu_{1}^{n}|^{2}=
p_{\mathrm{trans}}
\qquad W_{0}=p_{\mathrm{abs}}
\end{equation}

\subsection{Emission}

The initial state is
\begin{equation}\label{3.3.1}
|t=0\rangle=|\mathrm{M}0\rangle\otimes|\mathrm{Atom}1\rangle
\end{equation}
and the $\mathrm{U}$ evolution is
\begin{equation}\label{3.3.2}
|\mathrm{M}0\rangle\otimes|\mathrm{Atom}1\rangle
\stackrel{\mathrm{U}}{\rightarrow}\mu_{0}^{1}
|\mathrm{M}0\rangle\otimes|\mathrm{Atom}1\rangle+ \mu_{1}^{1}
|\mathrm{M}1\rangle\otimes|\mathrm{Atom}0\rangle
\end{equation}
so that
\begin{equation}\label{3.3.3}
\hat{\rho}_{\mathrm{red}}^{1}(t)=
\sum_{j}^{0,1}|\mu_{j}^{1}|^{2}|j\rangle\langle j|
\end{equation}
\begin{equation}\label{3.3.4}
|0\rangle=|\mathrm{M}0\rangle\otimes|\mathrm{Atom}1\rangle \qquad
|1\rangle=|\mathrm{M}1\rangle\otimes|\mathrm{Atom}0\rangle
\end{equation}
We have
\begin{equation}\label{3.3.5}
\mu_{0}^{1}(t)\rightarrow 0\qquad\mu_{1}^{1}(t)\rightarrow1\qquad
\mathrm{for}\;\;t\rightarrow\infty
\end{equation}
From (1.2.10), (1.2.11) follows
\begin{equation}\label{3.3.6}
W_{0}^{n}=w_{0}^{1}w_{0}^{2}\cdots
w_{0}^{n}=\left(\frac{1}{2}\right)^{n}\qquad
W_{1}^{n}=1-\left(\frac{1}{2}\right)^{n}\qquad n=1,2,\cdots
\end{equation}

Let the $\mathrm{U}$ evolution be such that
\begin{equation}\label{3.3.7}
|\mu_{0}^{1}|^{2}(t)=\mathrm{e}^{-t/\tau}
\end{equation}
Then
\begin{equation}\label{3.3.8}
\mathrm{e}^{-t^{1}_{\mathrm{red}}/\tau}=\frac{1}{2}\qquad
t^{1}_{\mathrm{red}}=(\ln 2)\tau
\end{equation}
and under the evolution with the reductions
\begin{equation}\label{3.1.9}
W_{0}(t)=\mathrm{e}^{-t/\tau}\qquad 0\leq t<\infty
\end{equation}

\subsection{Particle detection}

Consider the detection of a particle (electron, photon, atom) via
a (secondary) photon emission. The initial state is
\begin{equation}\label{3.4.1}
|t=0\rangle=\left[\bigotimes_{s=1}^{N}|\mathrm{M}_{s}0\rangle\right]
\otimes\left[\sum_{s=1}^{N}c_{s}|\mathrm{DPs1}\rangle\right]
\end{equation}
where $\mathrm{DP}$ stands for detector+particle. The unitary
evolution is this:
\begin{eqnarray}
|t=0\rangle\stackrel{\mathrm{U}}{\rightarrow}\sum_{s=1}^{N}c_{s}
\bigotimes_{s'}^{s'\neq s}|\mathrm{M}_{s'}0\rangle\otimes
\{\mu_{s0}^{1}|\mathrm{M}_{s}0\rangle\otimes|\mathrm{DP}s1\rangle
+\mu_{s1}^{1}|\mathrm{M}_{s}1\rangle\otimes|\mathrm{DP}s0\rangle\}\nonumber\\
=\left[\bigotimes_{s=1}^{N}|\mathrm{M}_{s}0\rangle\right]\otimes
\left[\sum_{s=1}^{N}c_{s}\mu_{s0}^{1}|\mathrm{DPs1}\rangle\right]
+\sum_{s=1}^{N}c_{s}\mu_{s1}^{1}\left[|\mathrm{M}_{s}1\rangle
\bigotimes_{s'}^{s'\neq
s}|\mathrm{M}_{s'}0\rangle\right]\otimes|\mathrm{DP}s0\rangle
\end{eqnarray}
Thus
\begin{equation}\label{3.4.3}
\hat{\rho}_{\mathrm{red}}^{1}(t)=\sum_{j=
0}^{N}w_{j}^{1}|j\rangle\langle j|
\end{equation}
with the states
\begin{equation}\label{3.4.4}
|0\rangle=\frac{1}{\sqrt{w_{0}^{1}}}\left[\bigotimes_{s=1}^{N}
|\mathrm{M}_{s}0\rangle\right]\otimes\left[\sum_{s=1}^{N}
c_{s}\mu_{s0}^{1}|\mathrm{DPs1}\rangle\right]\qquad|s\rangle=
\left[|\mathrm{M}_{s}1\rangle\bigotimes_{s'}^{s'\neq s}
|\mathrm{M}_{s'}0\rangle\right]\otimes|\mathrm{DPs0}\rangle
\end{equation}
and the probabilities
\begin{equation}\label{3.4.5}
w_{0}^{1}=\sum_{s=1}^{N}|c_{s}|^{2}|\mu_{s0}^{1}|^{2}\qquad
w_{s}^{1}=|c_{s}|^{2}|\mu_{s1}^{1}|^{2}
\end{equation}
The reduction is determined by (1.2.8):
\begin{equation}\label{3.4.6}
\sum_{j=0}^{N}(\ln
w_{j}^{1})\frac{\mathrm{d}w_{j}^{1}}{\mathrm{d}t}=0
\qquad\mathrm{with}\;\sum_{j=0}^{N}\frac{\mathrm{d}w_{j}^{1}}{\mathrm{d}t}=0
\end{equation}

Consider the simplest case:
\begin{equation}\label{3.4.7}
w_{1}^{1}=w_{2}^{1}=\cdots=w_{N}^{1}\qquad w_{0}^{1}=1-Nw_{1}^{1}
\qquad \frac{\mathrm{d}w_{0}^{1}}{\mathrm{d}t}=-N
\frac{\mathrm{d}w_{1}^{1}}{\mathrm{d}t}
\end{equation}
Now (3.4.6) boils down to
\begin{equation}\label{3.4.8}
[\ln(1-Nw_{1}^{1})]\left(-N\frac{\mathrm{d}w_{1}^{1}}{\mathrm{d}t}\right)+
N(\ln w_{1}^{1})\frac{\mathrm{d}w_{1}^{1}}{\mathrm{d}t}=0
\end{equation}
i.e., in view of $\mathrm{d}w_{1}^{1}/\mathrm{d}t\neq 0$,
\begin{equation}\label{3.4.9}
\ln\left(\frac{1}{w_{1}^{1}}-N\right)=0\qquad
w_{1}^{1}=\frac{1}{N+1}
\end{equation}
Thus
\begin{equation}\label{3.4.10}
w_{0}^{1}=w_{s}^{1}=\frac{1}{N+1}\,,\;\;\;s=1,2,\cdots,N
\end{equation}
After $n$ reductions
\begin{equation}\label{3.4.11}
W_{0}^{n}=\left(\frac{1}{N+1}\right)^{n}\qquad
W_{s}^{1}=\frac{1}{N}
\left[1-\left(\frac{1}{N+1}\right)^{n}\right]
\end{equation}

\subsection{The spectral line narrowing effect}

It is important to note the following. In the case of emission
from one source into one mode, $w_{0}^{1}=1/2$ (3.3.6); whereas in
the case of $N$ sources with related $N$ modes,
$w_{0}^{1}=1/(N+1)$. This results in the narrowing of a spectral
line with increasing $N$.

\subsection{Reduction of a superposition}

Consider the reduction of a superposition of two states via
interaction with a particle resulting in a photon emission. The
initial state is
\begin{equation}\label{3.6.1}
|t=0\rangle=|\mathrm{M}_{s}0\rangle\otimes\left[|\mathrm{P}s\rangle
\otimes\sum_{s'}^{1,2}c_{s'}|\mathrm{S}s'\rangle\right]
\end{equation}
where $\mathrm{P}$ stands for particle and $\mathrm{S}$ for
system. The unitary evolution is this:
\begin{eqnarray}
|t=0\rangle\stackrel{\mathrm{U}}{\rightarrow}
c_{\bar{s}}|\mathrm{M}_{s}0\rangle\otimes|\mathrm{P}s\rangle
\otimes|\mathrm{S}\bar{s}\rangle+c_{s}
\{\mu_{0}^{1}|\mathrm{M}_{s}0\rangle\otimes|\mathrm{P}s\rangle
\otimes|\mathrm{S}s\rangle+
\mu_{1}^{1}|\mathrm{M}_{s}1\rangle\otimes|\mathrm{SPs}\rangle\}\nonumber\\
=|\mathrm{M}_{s}0\rangle\otimes
[c_{\bar{s}}|\mathrm{P}s\rangle\otimes|\mathrm{S}\bar{s}\rangle+
c_{s}\mu_{0}^{1}|\mathrm{P}s\rangle\otimes|\mathrm{S}s\rangle] +
c_{s}\mu_{1}^{1}|\mathrm{M}_{s}1\rangle\otimes|\mathrm{SPs}\rangle
\end{eqnarray}
where $s=1,2\Leftrightarrow \bar{s}=2,1$. Thus
\begin{equation}\label{3.6.3}
\hat{\rho}_{\mathrm{red}}^{1}(t)=
\sum_{j}^{0,1}w_{j}^{1}|j\rangle\langle j|
\end{equation}
\begin{equation}\label{3.6.4}
|0\rangle=\frac{1}{\sqrt{w_{0}^{1}}}|\mathrm{M}_{s}0\rangle\otimes
[c_{\bar{s}}|\mathrm{P}s\rangle\otimes|\mathrm{S}\bar{s}\rangle+
c_{s}\mu_{0}^{1}|\mathrm{P}s\rangle\otimes|\mathrm{S}s\rangle]\qquad
|1\rangle=|\mathrm{M}_{s}1\rangle\otimes|\mathrm{SP}s\rangle
\end{equation}
\begin{equation}\label{3.6.5}
w_{0}^{1}=|c_{\bar{s}}|^{2}+|c_{s}|^{2}|\mu_{0}^{1}|^{2}\qquad
w_{1}^{1}=|c_{s}|^{2}|\mu_{1}^{1}|^{2}
\end{equation}

The subsequent treatment is based on (1.2.10), (1.2.11). If
$|c_{s}|^{2}\leq 1/2$, then
\begin{equation}\label{3.6.6}
|\mu_{1}^{1}|^{2}(t_\mathrm{red}^{1})=1\qquad
|\mu_{0}^{1}|^{2}(t_\mathrm{red}^{1})=0\qquad
w_{1}^{1}=|c_{s}|^{2}\qquad w_{0}^{1}=|c_{\bar{s}}|^{2}
\end{equation}
and there occurs only one reduction, $n_{\mathrm{max}}=1$. If
under the unitary evolution
$|\mu_{0}^{1}|^{2}=\mathrm{e}^{-t/\tau}$, then
$t_\mathrm{red}^{1}=\infty$. But due to fluctuations,
$t_\mathrm{red}^{1}<\infty$.

If $|c_{s}|^{2}>1/2$, then $n_{\mathrm{max}}>1$. In any case,
after the last reduction, the states are
\begin{equation}\label{3.6.7}
|s\rangle=|1\rangle=
|\mathrm{M}_{s}1\rangle\otimes|\mathrm{SP}s\rangle\qquad
|\bar{s}\rangle=|0\rangle=|\mathrm{M}_{s}0\rangle\otimes|\mathrm{P}s\rangle
\otimes|\mathrm{S}\bar{s}\rangle
\end{equation}
with the probabilities
\begin{equation}\label{3.6.8}
W_{s}=W_{1}=|c_{s}|^{2}\qquad W_{\bar{s}}=W_{0}=|c_{\bar{s}}|^{2}
\end{equation}

We have
\begin{equation}\label{3.6.9}
|\mu_{0}^{1}|^{2}(t_\mathrm{red}^{1})= \left\{
\begin{array}{lr}
0 \qquad\quad\;\mathrm{for}\;|c_{s}|^{2}\leq 1/2\\
 1-|c_{s}|^{2}\;\mathrm{for}\;\;|c_{s}|^{2}>1/2
\end{array}
\right\}<\frac{1}{2}
\end{equation}
so that the spectral line narrowing effect takes place.

\section {Applications: Nonintegral photon states}

\subsection{One-mode nonintegral states}

A familiar example of a nonintegral photon state is that of a
laser mode: a coherent state. Consider the simplest case of the
formation of one-mode nonintegral states. The initial state is
\begin{equation}\label{4.1.1}
|t=0\rangle=[\alpha_{0}^{0}|\mathrm{Atom}0\rangle+
\alpha_{1}^{0}|\mathrm{Atom}1\rangle]\otimes|\mathrm{M}0\rangle\qquad
\alpha_{0}^{0}\alpha_{1}^{0}\neq 0
\end{equation}
and the unitary evolution is this:
\begin{eqnarray}
|t=0\rangle\stackrel{\mathrm{U}}{\rightarrow}|t\rangle=
\alpha_{0}^{0}|\mathrm{Atom}0\rangle\otimes|\mathrm{M}0\rangle+
\alpha_{1}^{0}\{\mu_{0}^{0}|\mathrm{Atom}1\rangle\otimes|\mathrm{M}0\rangle
+\mu_{1}^{0}|\mathrm{Atom}0\rangle\otimes|\mathrm{M}1\rangle\}\nonumber\\
=[\alpha_{0}^{0}|\mathrm{Atom}0\rangle+
\alpha_{1}^{0}\mu_{0}^{0}|\mathrm{Atom}1\rangle]\otimes|\mathrm{M}0\rangle
+\alpha_{1}^{0}\mu_{1}^{0}|\mathrm{Atom}0\rangle\otimes|\mathrm{M}1\rangle\hspace{11mm}
\end{eqnarray}
This is not the Schmidt decomposition. The latter is of the form
\begin{equation}\label{4.1.3}
|t\rangle=\sum_{j_{1}}^{1,2}c_{j_{1}}^{1}
\left[\sum_{k}^{0,1}\alpha_{j_{1}k}^{1}|\mathrm{Atom}k\rangle\right]\otimes
\left[\sum_{n}^{0,1}\mu_{j_{1}n}^{1}|\mathrm{M}n\rangle\right]
\end{equation}
In the case of a unitary evolution, (4.1.3) is valid for all
$t\geq 0$ so that the photon mode remains entangled with the atom.

After the first reduction, the state is
\begin{equation}\label{4.1.4}
|j_{1}t_\mathrm{red}^{1}\rangle=\left[\sum_{k}^{0,1}\alpha_{j_{1}k}^{1}
|\mathrm{Atom}k\rangle\right]\otimes
\left[\sum_{n}^{0,1}\mu_{j_{1}n}^{1}|\mathrm{M}n\rangle\right]
\qquad
\mathrm{with\;probability}\;\;w_{j_{1}}^{1}=|c_{j_{1}}^{1}|^{2}
\end{equation}
Again
\begin{eqnarray}
|\mathrm{Atom}0\rangle\otimes|\mathrm{M}n\rangle
\stackrel{\mathrm{U}}{\rightarrow}
|\mathrm{Atom}0\rangle\otimes|\mathrm{M}n\rangle\hspace{41mm}\nonumber\\
|\mathrm{Atom}1\rangle\otimes|\mathrm{M}0\rangle
\stackrel{\mathrm{U}}{\rightarrow}
\mu_{0}^{0}|\mathrm{Atom}1\rangle\otimes|\mathrm{M}0\rangle+
\mu_{1}^{0}|\mathrm{Atom}0\rangle\otimes|\mathrm{M}1\rangle\\
|\mathrm{Atom}1\rangle\otimes|\mathrm{M}1\rangle
\stackrel{\mathrm{U}}{\rightarrow}
\mu_{1}^{0}|\mathrm{Atom}1\rangle\otimes|\mathrm{M}1\rangle+
\mu_{2}^{0}|\mathrm{Atom}0\rangle\otimes|\mathrm{M}2\rangle\nonumber
\end{eqnarray}
so that going over to the Schmidt decomposition we obtain
\begin{equation}\label{4.1.6}
|j_{1}t_\mathrm{red}^{1}\rangle \stackrel{\mathrm{U}}{\rightarrow}
|j_{1}t>t_\mathrm{red}^{1}\rangle=\sum_{j_{2}}^{1,2}c_{j_{1}j_{2}}^{2}
\left[\sum_{k}^{0,1}\alpha_{j_{1}j_{2}k}^{2}|\mathrm{Atom}k\rangle\right]\otimes
\left[\sum_{n=0}^{2}\mu_{j_{1}j_{2}n}^{2}|\mathrm{M}n\rangle\right]
\end{equation}
After the second reduction,
\begin{equation}\label{4.1.7}
|j_{1}j_{2}t_\mathrm{red}^{2}\rangle=
\left[\sum_{k}^{0,1}\alpha_{j_{1}j_{2}k}^{2}|\mathrm{Atom}k\rangle\right]\otimes
\left[\sum_{n=0}^{2}\mu_{j_{1}j_{2}n}^{2}|\mathrm{M}n\rangle\right]
\end{equation}
with the conditional and total probabilities
\begin{equation}\label{4.1.8}
w_{j_{1}j_{2}}^{2}=|c_{j_{1}j_{2}}^{2}|^{2}\qquad
W_{j_{1}j_{2}}^{2}= w_{j_{1}}^{1}w_{j_{1}j_{2}}^{2}
\end{equation}
After the $r$-th reduction,
\begin{equation}\label{4.1.9}
|j_{1}j_{2}\cdots j_{r}t_{\mathrm{red}}^{r}\rangle=
\left[\sum_{k}^{0,1}\alpha_{j_{1}j_{2}\cdots j_{r}
k}^{r}|\mathrm{Atom}k\rangle\right]\otimes
\left[\sum_{n=0}^{r}\mu_{j_{1}j_{2}\cdots j_{r}
n}^{r}|\mathrm{M}n\rangle\right]
\end{equation}
with the probabilities
\begin{equation}\label{4.1.10}
w_{j_{1}j_{2}\cdots j_{r}}^{r}=|c_{j_{1}j_{2}\cdots
j_{r}}^{r}|^{2}\qquad W_{j_{1}j_{2}\cdots j_{r}}^{r}=
w_{j_{1}}^{1}w_{j_{1}j_{2}}^{2}\cdots w_{j_{1}j_{2}\cdots
j_{r}}^{r}
\end{equation}

\subsection{Entangled pair of photons}

Consider the reduction of an entangled pair of photons via
absorption. The initial state is
\begin{equation}\label{4.2.1}
|t=0\rangle=\left[\sum_{l}^{1,2}c_{l}|\mathrm{M}al\rangle\otimes
|\mathrm{M}b\bar{l}\rangle\right]\otimes|\mathrm{R}0\rangle \qquad
l=1,2\Leftrightarrow \bar{l}=2,1
\end{equation}
where the energy related to the mode $ |\mathrm{M}a/bl\rangle$
with the location $a/b$ is $\omega_{l}$. The unitary evolution is
of the form
\begin{eqnarray}
|\mathrm{M}al\rangle\otimes
|\mathrm{M}b\bar{l}\rangle\otimes|\mathrm{R}0\rangle
\stackrel{\mathrm{U}}{\rightarrow}\nonumber\\ \mu_{alb\bar{l}}
|\mathrm{M}al\rangle\otimes|\mathrm{M}b\bar{l}\rangle \otimes
|\mathrm{R}0\rangle\nonumber\\ +\mu_{al}^{1}|\mathrm{M}al\rangle
\otimes|\mathrm{M}b0\rangle \otimes
|\mathrm{R}b\bar{l}\rangle\nonumber\\ +
\mu_{b\bar{l}}^{1}|\mathrm{M}a0\rangle
\otimes|\mathrm{M}b\bar{l}\rangle \otimes
|\mathrm{R}al\rangle\nonumber\\ +\mu_{l}^{1}|\mathrm{M}a0\rangle
\otimes|\mathrm{M}b0\rangle \otimes |\mathrm{R}alb\bar{l}\rangle
\end{eqnarray}
so that the Schmidt decomposition is this:
\begin{eqnarray}
|t=0\rangle\stackrel{\mathrm{U}}{\rightarrow}|t\rangle\hspace{40mm}\nonumber\\
=\left[\sum_{l}^{1,2}c_{l}\mu_{alb\bar{l}}
|\mathrm{M}al\rangle\otimes|\mathrm{M}b\bar{l}\rangle\right]
\otimes|\mathrm{R}0\rangle\nonumber\\
+\sum_{l}^{1,2}c_{l}\mu_{al}^{1}
[\,|\mathrm{M}al\rangle\otimes|\mathrm{M}b0\rangle\,]
\otimes|\mathrm{R}b\bar{l}\rangle\nonumber\\
+\sum_{l}^{1,2}c_{l}\mu_{b\bar{l}}^{1}
[\,|\mathrm{M}a0\rangle\otimes|\mathrm{M}b\bar{l}\rangle\,]
\otimes|\mathrm{R}al\rangle\nonumber\\ +[\,|\mathrm{M}a0\rangle
\otimes|\mathrm{M}b0\rangle\,]\otimes\left[\sum_{l}^{1,2}c_{l}\mu_{l}^{1}
|\mathrm{R}alb\bar{l}\rangle\right]
\end{eqnarray}
The states after the first reduction are:
\begin{equation}\label{4.2.4}
|2t_{\mathrm{red}}^{1}\rangle:=\frac{1}{\sqrt{w_{2}}}
\left[\sum_{l}^{1,2}c_{l}\mu_{alb\bar{l}}
|\mathrm{M}al\rangle\otimes|\mathrm{M}b\bar{l}\rangle\right]
\otimes|\mathrm{R}0\rangle
\end{equation}
\begin{equation}\label{4.2.5}
|alt_{\mathrm{red}}^{1}\rangle:=
[\,|\mathrm{M}al\rangle\otimes|\mathrm{M}b0\rangle\,]
\otimes|\mathrm{R}b\bar{l}\rangle\,,\;l=1,2
\end{equation}
\begin{equation}\label{4.2.6}
|blt_{\mathrm{red}}^{1}\rangle:=
[\,|\mathrm{M}a0\rangle\otimes|\mathrm{M}bl\rangle\,]
\otimes|\mathrm{R}a\bar{l}\rangle\,,\;l=1,2
\end{equation}
\begin{equation}\label{4.2.7}
|0t_{\mathrm{red}}^{1}\rangle:=\frac{1}{\sqrt{w_{0}}}
[\,|\mathrm{M}a0\rangle \otimes|\mathrm{M}b0\rangle\,]\otimes
\left[\sum_{l}^{1,2}c_{l}\mu_{l}^{1}
|\mathrm{R}alb\bar{l}\rangle\right]
\end{equation}
with the probabilities
\begin{equation}\label{4.2.8}
w_{2}=\sum_{l}^{1,2}|c_{l}|^{2}|\mu_{alb\bar{l}}^{1}|^{2}\qquad
w_{al}=|c_{l}|^{2}|\mu_{al}^{1}|^{2}\qquad
w_{bl}=|c_{\bar{l}}|^{2}|\mu_{bl}^{1}|^{2}\qquad
w_{0}=\sum_{l}^{1,2}|c_{l}|^{2}|\mu_{l}^{1}|^{2}
\end{equation}
The subsequent treatment presents no special problems.

\subsection{Atom-photon entanglement}

Consider the reduction of an atom-photon entanglement due to
photon absorption. The initial state is
\begin{equation}\label{4.3.1}
|t=0\rangle=\sum_{l}^{1,2}c_{l}[\,|\mathrm{M_{l}1}\rangle\otimes
|\mathrm{M}_{\bar{l}}0\rangle\,]\otimes[\,|\mathrm{Atom}l\rangle
\otimes|\mathrm{R}0\rangle\,]\qquad l=1,2\Leftrightarrow
\bar{l}=2,1
\end{equation}
We have
\begin{equation}\label{4.3.2}
|\mathrm{M}_{l}1\rangle\otimes|\mathrm{R}0\rangle
\stackrel{\mathrm{U}}{\rightarrow}\mu_{l1}^{1}
|\mathrm{M}_{l}1\rangle\otimes|\mathrm{R}0\rangle+
\mu_{l0}^{1}|\mathrm{M}_{l}0\rangle\otimes|\mathrm{R}l\rangle
\end{equation}
so that
\begin{eqnarray}
|t=0\rangle\stackrel{\mathrm{U}}{\rightarrow} |t\rangle=
\sum_{l}^{1,2}c_{l}\mu_{l1}^{1}[\,|\mathrm{M_{l}1}\rangle\otimes
|\mathrm{M}_{\bar{l}}0\rangle\,]\otimes[\,|\mathrm{Atom}l\rangle
\otimes|\mathrm{R}0\rangle\,]\nonumber\\
\qquad\qquad\qquad+[|\mathrm{M}_{1}0\rangle\otimes|\mathrm{M}_{2}0\rangle]
\otimes\left[\sum_{l}^{1,2}c_{l}\mu_{l0}^{1}|\mathrm{Atom}l\rangle
\otimes|\mathrm{R}l\rangle\right]
\end{eqnarray}
Both (4.3.1) and (4.3.3) are the Schmidt decompositions.

The states after the first reduction are:
\begin{equation}\label{4.3.4}
|lt_{\mathrm{red}}^{1}\rangle:=[\,|\mathrm{M_{l}1}\rangle\otimes
|\mathrm{M}_{\bar{l}}0\rangle\,]\otimes[\,|\mathrm{Atom}l\rangle
\otimes|\mathrm{R}0\rangle\,]\qquad
w_{l}=|c_{l}|^{2}|\mu_{l1}^{1}|^{2}\,,\;\,l=1,2
\end{equation}
\begin{equation}\label{4.3.5}
|0t_{\mathrm{red}}^{1}\rangle:=\frac{1}{\sqrt{w_{0}}}
[\,|\mathrm{M}_{1}0\rangle\otimes|\mathrm{M}_{2}0\rangle\,]
\otimes\left[\sum_{l}^{1,2}c_{l}\mu_{l0}^{1}|\mathrm{Atom}l\rangle
\otimes|\mathrm{R}l\rangle\right]\qquad
w_{0}=\sum_{l}^{1,2}|c_{l}|^{2}|\mu_{l0}^{1}|^{2}
\end{equation}

We have
\begin{equation}\label{4.3.6}
\sigma(t=0)=-\sum_{l}^{1,2}|c_{l}|^{2}\ln|c_{l}|^{2}
\end{equation}
and
\begin{equation}\label{4.3.7}
\sigma(t>0)=-\sum_{l}^{1,2}w_{l}\ln w_{l}-w_{0}\ln w_{0}
\end{equation}
Let
\begin{equation}\label{4.3.8}
|\mu_{11}^{1}|^{2}=|\mu_{21}^{1}|^{2}=:|\mu_{1}^{1}|^{2}\qquad
|\mu_{10}^{1}|^{2}=|\mu_{20}^{1}|^{2}=:|\mu_{0}^{1}|^{2}\qquad
|\mu_{1}^{1}|^{2}+|\mu_{0}^{1}|^{2}=1
\end{equation}
then
\begin{equation}\label{4.3.9}
\sigma(t>0)=(1-|\mu_{0}^{1}|^{2})\sigma(t=0)+
(-|\mu_{1}^{1}|^{2}\ln|\mu_{1}^{1}|^{2}
-|\mu_{0}^{1}|^{2}\ln|\mu_{0}^{1}|^{2})
\end{equation}
From
\begin{equation}\label{4.3.10}
\frac{\mathrm{d}[\sigma(t>0)]}{\mathrm{d}t}=0
\end{equation}
follows
\begin{equation}\label{4.3.11}
|\mu_{0}^{1}|^{2}(t_{\mathrm{red}}^{1})=
\frac{1}{1+\mathrm{e}^{\sigma(t>0)}}\qquad
|\mu_{1}^{1}|^{2}(t_{\mathrm{red}}^{1})=
\frac{1}{1+\mathrm{e}^{-\sigma(t>0)}}
\end{equation}
Next,
\begin{equation}\label{4.3.12}
|lt_{\mathrm{red}}^{1}\rangle\stackrel{\mathrm{U}}{\rightarrow}
\mu_{1}^{1}[\,|\mathrm{M_{l}1}\rangle\otimes
|\mathrm{M}_{\bar{l}}0\rangle\,]\otimes[\,|\mathrm{Atom}l\rangle
\otimes|\mathrm{R}0\rangle\,]+
\mu_{0}^{1}[\,|\mathrm{M}_{1}0\rangle\otimes|\mathrm{M}_{2}0\rangle\,]
\otimes[\,|\mathrm{Atom}l\rangle \otimes|\mathrm{R}l\rangle\,]
\end{equation}
so that after the second reduction
\begin{equation}\label{4.3.13}
|\mu_{1}^{1}|^{2}(t_{\mathrm{red}}^{2})=
|\mu_{0}^{1}|^{2}(t_{\mathrm{red}}^{2})=\frac{1}{2}
\end{equation}
The subsequent treatment is trivial.

\section{Applications: Weak bosons and gluons}

\subsection{A process with a weak boson}

Now let us consider a process with an intermediate weak boson. The
initial state is
\begin{equation}\label{5.1.1}
|t=0\rangle=|\mathrm{in}\rangle=
|\mathrm{P_{in}}\rangle\otimes|\mathrm{W}0\rangle
\end{equation}
where $\mathrm{P}$ stands for particle and $\mathrm{W}$ for weak
boson mode. The unitary evolution is this:
\begin{equation}\label{5.1.2}
|t=0\rangle=|\mathrm{in}\rangle\stackrel{\mathrm{U}}{\rightarrow}
|t\rangle=\alpha_{\mathrm{in}}|\mathrm{in}\rangle+
\alpha_{\mathrm{inter}}|\mathrm{inter}\rangle+
\alpha_{\mathrm{out}}|\mathrm{out}\rangle
\end{equation}
where
\begin{equation}\label{5.1.3}
|\mathrm{inter}\rangle=|\mathrm{P_{out}}_{1}\rangle\otimes
|\mathrm{W}1\rangle\qquad|\mathrm{out}\rangle=
|\mathrm{P_{out}}_{1}\rangle\otimes|\mathrm{P_{out}}_{2}\rangle
\otimes|\mathrm{P_{out}}_{3}\rangle\otimes|\mathrm{W}0\rangle
\end{equation}
The Schmidt decomposition is
\begin{equation}\label{5.1.4}
|t\rangle=[\alpha_{\mathrm{in}}|\mathrm{in}\rangle+
\alpha_{\mathrm{out}}|\mathrm{P_{\mathrm{out}}}_{123}\rangle]
\otimes|\mathrm{W}0\rangle+
\alpha_{\mathrm{inter}}|\mathrm{P_{out}}_{1}\rangle\otimes
|\mathrm{W}1\rangle
\end{equation}
where
\begin{equation}\label{5.1.5}
|\mathrm{P_{\mathrm{out}}}_{123}\rangle:=
|\mathrm{P_{out}}_{1}\rangle\otimes|\mathrm{P_{out}}_{2}\rangle
\otimes|\mathrm{P_{out}}_{3}\rangle
\end{equation}
The states after the first reduction are these:
\begin{equation}\label{5.1.6}
|0t_{\mathrm{red}}^{1}\rangle=\frac{1}{\sqrt{w_{0}}}
[\alpha_{\mathrm{in}}|\mathrm{in}\rangle+
\alpha_{\mathrm{out}}|\mathrm{P_{\mathrm{out}}}_{123}\rangle]
\otimes|\mathrm{W}0\rangle\qquad
w_{0}=|\alpha_{\mathrm{in}}|^{2}+|\alpha_{\mathrm{out}}|^{2}=:
w_{\mathrm{in}}+w_{\mathrm{out}}
\end{equation}
\begin{equation}\label{5.1.7}
|1t_{\mathrm{red}}^{1}\rangle=|\mathrm{P_{out}}_{1}\rangle\otimes
|\mathrm{W}1\rangle\qquad w_{1}=|\alpha_{\mathrm{inter}}|^{2}
\end{equation}
The entropy is
\begin{equation}\label{5.1.8}
\sigma(t)=-w_{0}\ln w_{0}-w_{1}\ln w_{1}=-(1-w_{1})\ln(1-w_{1})
\end{equation}
and
\begin{equation}\label{5.1.9}
\frac{\mathrm{d}\sigma}{\mathrm{d}t}=
\left[\ln\left(\frac{1}{w_{1}}-1\right)\right]
\frac{\mathrm{d}w_{1}}{\mathrm{d}t}
\end{equation}
so that either
\begin{equation}\label{5.1.10}
w_{1}=\frac{1}{2}
\end{equation}
or
\begin{equation}\label{5.1.11}
\frac{\mathrm{d}w_{1}}{\mathrm{d}t}=0\qquad w_{1}\leq \frac{1}{2}
\end{equation}

In the case
$|t_{\mathrm{red}}^{1}\rangle=|0t_{\mathrm{red}}^{1}\rangle$ the
unitary evolution is determined by (5.1.2), and in the case
$|t_{\mathrm{red}}^{1}\rangle=|1t_{\mathrm{red}}^{1}\rangle$ by
\begin{equation}\label{5.1.12}
|\mathrm{W}1\rangle\stackrel{\mathrm{U}}{\rightarrow}
\alpha_{1}|\mathrm{W}1\rangle+\alpha_{0}
|\mathrm{P_{\mathrm{out}}}_{123}\rangle\otimes|\mathrm{W}0\rangle
\end{equation}
which is the Schmidt decomposition. The subsequent treatment is
standard.

It should be particularly emphasized that it is the reduction that
disentangles the weak boson.

Let under the unitary evolution
\begin{equation}\label{5.1.13}
\mathrm{d}w_{\mathrm{in}}=
-\lambda_{\mathrm{in}}w_{\mathrm{in}}\mathrm{d}t\qquad
\mathrm{d}w_{1}=\lambda_{\mathrm{in}}w_{\mathrm{in}}\mathrm{d}t-
\lambda_{1}w_{1}\mathrm{d}t
\end{equation}
whence
\begin{equation}\label{5.1.14}
w_{\mathrm{in}}=\mathrm{e}^{-\lambda_{\mathrm{in}}t}\qquad w_{1}=
\frac{\lambda_{\mathrm{in}}}{\lambda_{1}-\lambda_{\mathrm{in}}}
\left(\mathrm{e}^{-\lambda_{\mathrm{in}}t}
-\mathrm{e}^{-\lambda_{1}t}\right)
\end{equation}
and
\begin{equation}\label{5.1.15}
w_{\mathrm{out}}=1-w_{\mathrm{in}}-w_{1}=
1-\frac{1}{\lambda_{1}-\lambda_{\mathrm{in}}}
\left(\lambda_{1}\mathrm{e}^{-\lambda_{\mathrm{in}}t}
-\lambda_{\mathrm{in}}\mathrm{e}^{-\lambda_{1}t}\right)
\end{equation}
Now $\mathrm{d}w_{1}/\mathrm{d}t=0$ results in
\begin{equation}\label{5.1.16}
t=t_{0}:=\frac{\ln(\lambda_{1}/\lambda_{\mathrm{in}})}
{\lambda_{\mathrm{in}}[(\lambda_{1}/\lambda_{\mathrm{in}})-1]}
\end{equation}
Introduce
\begin{equation}\label{5.1.17}
\tau=\lambda_{\mathrm{in}}\qquad
\beta=\lambda_{1}/\lambda_{\mathrm{in}}
\end{equation}
then
\begin{equation}\label{5.1.18}
w_{\mathrm{in}}=\mathrm{e}^{-\tau}\qquad
w_{1}=\frac{\mathrm{e}^{-\tau}}{\beta-1}
\left[1-\mathrm{e}^{-(\beta-1)\tau}\right]\qquad
w_{\mathrm{out}}=1-\frac{1}{\beta-1}
\left(\beta\mathrm{e}^{-\tau}-\mathrm{e}^{-\beta\tau}\right)
\end{equation}
\begin{equation}\label{5.1.19}
\tau_{0}=\frac{\ln \beta}{\beta-1}
\end{equation}
and
\begin{equation}\label{5.1.20}
w_{1}(\tau_{0})=
\frac{1}{\beta}\mathrm{e}^{\ln \beta/(1-\beta)}
\end{equation}

Specifically,
\begin{eqnarray}
\hspace{-17mm}\mathrm{for}\;\beta\gg 1\quad
w_{1}(\tau_{0})\approx\frac{1}{\beta}\ll
1\quad\tau_{\mathrm{red}}^{1}=\tau_{0}=\frac{\ln
\beta}{\beta}\hspace{14mm} \nonumber\\ \mathrm{for}\;\beta\ll
1\quad w_{1}(\tau_{0})\approx 1>\frac{1}{2}\quad
w_{1}(\tau_{\mathrm{red}}^{1})=
\frac{1}{2}\quad\tau_{\mathrm{red}}^{1}\approx \ln 2
\\ \mathrm{for}\;\beta=1\quad
w_{1}(\tau_{0})=\frac{1}{\mathrm{e}}<
\frac{1}{2}\quad\tau_{\mathrm{red}}^{1}=\tau_{0}=1\hspace{22mm}\nonumber
\end{eqnarray}

\subsection{On the role of gluons}

Finally, let us dwell on the reduction problem in nuclear decay
(specifically, alpha decay). In such a problem, gauge bosons are
gluons. But as long as there are no free gluons, the only
reasonable conclusion is this: A reduction in a nuclear decay is
due to a change of the state of gluon degrees of freedom.

\section*{Acknowledgments}

I would like to thank Alex A. Lisyansky for support and Stefan V.
Mashkevich for helpful discussions.

\end{document}